\documentclass[journal]{IEEEtran}
\usepackage{amsmath,amsfonts}
\usepackage{algorithmic}
\usepackage{algorithm}
\usepackage{array}
\usepackage[caption=false,font=normalsize,labelfont=sf,textfont=sf]{subfig}
\usepackage{textcomp}
\usepackage{stfloats}
\usepackage{url}
\usepackage{verbatim}
\usepackage{graphicx}
\usepackage{cite}
\usepackage{color}
\usepackage{amssymb}

\begin{document}

\title{Joint Precoding for RIS-Assisted Wideband THz Cell-Free Massive MIMO Systems}


\author{Xin~Su,~\IEEEmembership{Student Member,~IEEE}, Ruisi~He,~\IEEEmembership{Senior Member,~IEEE}, Peng~Zhang, and Bo~Ai,~\IEEEmembership{Fellow,~IEEE}  

\thanks{X. Su, R. He, P. Zhang and B. Ai are with the School of Electronics and Information Engineering, Beijing Jiaotong University, Beijing 100044, China (e-mail: 22120124@bjtu.edu.cn; ruisi.he@bjtu.edu.cn; jumper@bjtu.edu.cn; boai@bjtu.edu.cn).}
}

\maketitle
\begin{abstract}
Terahertz (THz) cell-free massive multiple-input-multiple-output (mMIMO) networks have been envisioned as a prospective technology for achieving higher system capacity, improved performance, and ultra-high reliability in 6G networks. However, due to severe attenuation and limited scattering in THz transmission, as well as high power consumption for increased number of access points (APs), further improvement of network capacity becomes challenging. Reconfigurable intelligent surface (RIS) has been introduced as a low-cost solution to reduce AP deployment and assist in data transmission. However, due to the ultra-wide bandwidth and frequency-dependent characteristics of RISs, beam split effect has become an unavoidable obstacle.
To compensate the severe performance degradation caused by beam split effect, we introduce additional time delay (TD) layers at both access points (APs) and RISs. Accordingly, we propose a joint precoding framework at APs and RISs to fully unleash the potential of the considered network. Specifically, we first formulate the joint precoding as a non-convex optimization problem. Then, given the location of unchanged RISs, we adjust the time delays (TDs) of APs to align the generated beams towards RISs. After that, with knowledge of the optimal TDs of APs, we decouple the optimization problem into three subproblems of optimizing the baseband beamformers, RISs and TDs of RISs, respectively. Exploiting multidimensional complex quadratic transform, we transform the subproblems into convex forms and solve them under alternate optimizing framework. Numerical results verify that the proposed method can effectively mitigate beam split effect and significantly improve the achievable rate compared with conventional cell-free mMIMO networks.

\end{abstract}


\section{Introduction}
\IEEEPARstart{T}{he} forthcoming 6G is expected to provide unparalleled data speeds, ultra-low latency, and seamless integration of advanced technologies to support emerging applications spanning from
industrial Internet of Things (IoT) to intelligent railway and other new-generation systems \cite{rose2015internet,he20225g,huang2022artificial}, which presents a hundredfold increase in demands for system capacity compared to 5G \cite{chen2020vision,he2019applications}. However, the enhancement of network capacity in conventional cellular networks is constrained by inter-cell interference \cite{ngo2017cell}. Fortunately, the recently proposed network paradigm cell-free massive multiple-input multiple-output (mMIMO) is envisioned as a promising technology to break the above constraints \cite{andrews2016we}. Unlike traditional cellular networks, cell-free mMIMO systems deploy distributed access points (APs) that cooperate without cell boundaries to effectively mitigate the inter-cell interference while enhancing system capacity \cite{bjornson2020scalable}. To further improve the capacity, more distributed APs should be deployed in cell-free mMIMO systems, which requires high power consumption and unaffordable cost. Thanks to the introduction of reconfigurable intelligent surface (RIS), which is considered to be a counterpart for enhancing channel capacity and improving network efficiency \cite{lan2023new}. Equipped with a large number of high-gain passive elements, RIS can intelligently manipulate the elertromagentic environment without additional energy consumption \cite{bjornson2022reconfigurable}. Therefore, the integration of RIS into cell-free-mMIMO networks holds significant potential for optimizing network capacity and further improving the overall system performance \cite{shi2022wireless}. 

Apart from the employment of distributed architecture, the expansion of spectrum resources is of great importance for the improvement of network capacity. Terahertz (THz) band (0.1-10 THz), which is capable of providing tens of GHz unlicensed bandwidth, is considered to support increased transmission rates \cite{he2019propagation,song2011present}. However, due to the susceptibility to blockage and limited scatterings, THz signals highly depend on line-of-sight (LoS) propagation. To address this issue, RIS can be thankfully exploited once again to provide virtual LoS links for THz communications \cite{du2022performance}. 

As discussed above, RIS-assisted THz cell-free mMIMO wireless systems have shown prospective potential to satisfy the demand of future 6G networks. This innovative paradigm can not only enhance system capacity but also significantly improve the overall network performance. 

\subsection{Prior Works}
To fully unleash the potential of RISs in cell-free mMIMO systems, the joint design of active precoding at APs and passive beamforming at RISs is the key technique to guarantee the optimal network capacity and achievable rate \cite{zhang2021joint,zhang2023joint,ma2023cooperative,jin2022ris}. Specifically, the joint precoding framework of RIS-aided cell-free framework is firstly proposed in \cite{zhang2021joint}. Besides, the authors in \cite{zhang2023joint} proposed a novel joint distributed precoding and beamforming method by decentralizing the alternating optimization (AO) problem to minimize the weighted sum of users’ mean square error. Then, the cooperative beamforming is formulated as a non-convex weighted sum-rate (WSR) maximization problem and solved by fractional programming algorithm in \cite{ma2023cooperative}. Moreover, in \cite{jin2022ris}, the joint beamforming is transformed into as a max–min fairness problem.   

The above solutions have significantly improved the WSR performance in RIS-assisted cell-free networks. However, since hybrid beamforming architecture is typically employed for THz mMIMO as a performance-cost trade-off, these schemes might become inapplicable for THz communication systems \cite{zhang2015massive}. Consequently, hybrid precoding and beamforming in THz mMIMO communication systems is investigated in \cite{busari2019generalized,yan2020dynamic}. Specifically, a novel generalized framework for the design and performance analysis of the hybrid precoding architecture is presented in \cite{busari2019generalized}. The authors in \cite{yan2020dynamic} proposed a dynamic array-of-subarrays hybrid precoding
architecture to improve the data rate. 
However, for wideband THz systems, the deployment of hybrid beamforming architecture causes severe beam split effect. The generated beams at different subcarrier (SC) frequencies will split into distinct physical directions, resulting in severe array gain loss \cite{tan2021wideband}. To deal with this issue, the authors in \cite{dai2022delay} proposed a novel delay phase precoding architecture to mitigate beam split effect. In \cite{ning2023beamforming}, an effective beamforming strategy is proposed for multiuser THz ultra mMIMO. 

However, RIS also faces beam split effect due to its frequency-independent reflecting elements \cite{su2024channel}, which might present challenges for the effectiveness of the aforementioned methods. Currently, several works have investigated the mitigation of beam split effect in RIS-assisted THz systems \cite{su2023wideband,zhao2023joint,sun2023beamforming}. One hardware solution is to introduce additional time-delay (TD) modules and phase shifters into RIS elements to convert the phase-only precoding to joint phase and delay precoding \cite{su2023wideband}. Besides, the authors in \cite{zhao2023joint} applied TDs under practical constraints. Based on this, a double-layer true TD scheme is proposed in \cite{sun2023beamforming} for the distributed RISs-aided THz Communications. Furthermore, the joint beamforming design is investigated by employing deep reinforcement learning \cite{huang2021multi}. Moreover, the authors in \cite{mehrabian2024joint} developed a metapath-based heterogeneous graph-transformer network to avoid the performance loss caused by estimation inaccuracy.

In conclusion, there are numerous works dedicated in RIS-assisted cell-free mMIMO networks and RIS-assisted THz systems, respectively. However, to the best of our knowledge, there are no research investigating RIS-assisted THz cell-free mMIMO system with beam split effect, and the corresponding precoding problems have not been addressed well.

\subsection{Main Contributions}
Motivated by the above works, in this paper we investigate RIS-assisted THz cell-free mMIMO scenarios. We construct a three-dimensions framework, which includes bandwidth expansion, 
hardware deployment and precoding design, to enhance the system capacity and improve the overall performance with low cost and power consumption. The main contributions of this paper are summarized as follows:
\begin{itemize}
	\item To achieve capacity improvement, we investigate the downlink transmission in a practical RIS-assisted wideband THz cell-free mMIMO system where beam split effect is under consideration. We introduce an additional time delay (TD) layer at both APs and RISs to overcome the severe performance degradation caused by beam split effect. In addition to the employment of ideal RIS, we also discuss more practical low-resolution RIS cases where the phase shifts are discrete.
	\item We propose a novel joint precoding design for the considered scenario. Specifically, we derive the signal-to-interference-plus-noise ratio (SINR) and formulate the joint precoding problem as a WSR maximization optimization. Based on fact that the locations of RISs remain unchanged, we first optimize the TDs at APs to align the generated beams with  RISs at all SCs. Given the optimal TDs of APs, we then develop an AO framework and decouple the original WSR maximization problem into three subproblems, i.e., optimization of baseband beamformers, phase shifts and TDs of RISs. To solve the non-convex subproblems, we exploit multidimensional complex quadratic transform (MCQT) by introducing auxiliary variables to reformulate them into convex form.
	\item Numerical results demonstrate that the performance of the proposed joint precoding framework even equipped with low-resolution RISs can efficiently mitigate beam split effect and significantly enhance the network capacity. This indicates that the proposed framework can improve the network performance with low-cost and power consumption.
\end{itemize}

\subsection{Organization and Notations}
{\it{Organization}}: The remainder of this paper is organized as follows. In Sec. \ref{sec:System Model}, RIS-assisted THz cell-free mMIMO system model is formulated. The propsed optimization framework is introduced in Sec. \ref{sec:Proposed Method}. Numerical simulations are discussed in Sec. \ref{sec:Numerical Results and Discussion} and conclusions are drawn in Sec. \ref{sec:Conclusion}.  

{\it{Notations}}: Boldface lower-case and capital letters represent column vectors and matrices, respectively. $\left(\cdot \right)^*$, $\left(\cdot \right)^{\mathsf{T}}$ and $\left(\cdot \right)^{\mathsf{H}}$ denote conjugate, transpose and transpose-conjugate operation, respectively. $\mathbb{R}$, $\mathbb{R}^{+}$ and $\mathbb{C}$ denote the set of real, positive real and complex numbers, respectively. $\mathbf{I}_{N}$ is $N\times N$ size identity matrix and $\boldsymbol{0}_N$ is the $N$-length zero vector. $\boldsymbol{e}_l$ indicates elementary vector with a one at the $l$-th position, and $\boldsymbol{1}_L$ is an $L$-length vector with all elements equal 1. $\mathfrak{Re}$ denotes the real part of its argument, $|\cdot|$ and $\|\cdot\|$ denote the modulus and Euclidean norm of its argument, respectively. $\mathrm{diag}\{\cdot\}$, $\mathrm{coldiag}\{\cdot\}$ and $\mathrm{blkdiag}\{\cdot\}$ denote the diagonal operation, column diagonal operation, and generalized block diagonal operation, respectively.

\section{System Model}\label{sec:System Model}
In this section, we first introduce the hybrid precoding architecture with an additional TD layer. Then, we represent the RIS-assisted THz cell-free mMIMO system model and beam-split-affected channel model. Finally, we formulate the WSR maximization problem in the considered system.
\begin{figure}[!t]
	\centering
	\includegraphics[width=0.5\textwidth]{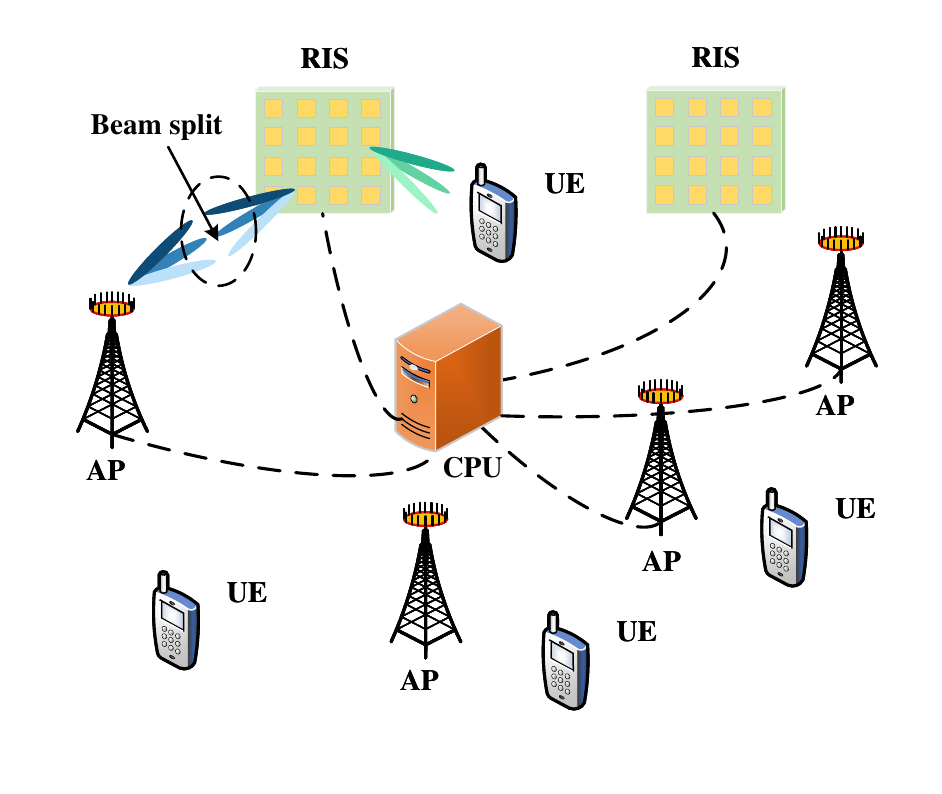}
	\caption{RIS-assisted THz cell-free mMIMO communication system with beam split.}
	\label{fig:System_Model}
\end{figure}
\subsection{Precoding Architecture}
We consider a RIS-assisted wideband THz cell-free mMIMO system as depicted in Fig. \ref{fig:System_Model}, where $A$ distributed APs and $R$ distributed RISs are deployed to cooperatively serve $K$ single-antenna users. Each AP is a uniform linear array (ULA) equipped with $N_{\mathrm{Tx}}$ antennas and $N_{\mathrm{RF}}$ RF-chains. For covenience, we assume $N_{\mathrm{RF}}=R$ in this paper, i.e., each RIS is served by a single RF-chain \cite{gao2016near}. Besides, Each RIS is a uniform planar array (UPA) equipped with $N_{\mathrm{RIS}}=N_{\mathrm{X}}\times N_{\mathrm{Y}}$ reflecting elements, where $N_{\mathrm{X}}$ and $N_{\mathrm{Y}}$ denote the numbers of elements in horizontal and vertical directions, respectively. Additionally, the OFDM transmission is employed where the bandwidth $B$ is divided into $M$ subcarriers (SCs). Without loss of generality, we assume that the perfect CSI is available via existing channel estimation algorithms \cite{su2024channel}. For expression simplification, we define $\mathcal{A}=\{1, \dots, A\}$, $\mathcal{R}=\{1, \dots, R\}$, $\mathcal{K}=\{1, \dots, K\}$ and $\mathcal{M}=\{1, \dots, M\}$   
as the index sets of the APs, RISs, UEs and SCs, respectively.

To deal with beam split effect, we introduce a hybrid precoding architecture with an additional TD layer between the RF chains and the analog beamformer. Specifically, each RF chain is connected to $D$ frequency-dependent TD elements and each TD element is connected to $N_{\mathrm{D}}=N_{\mathrm{Tx}}/D$ frequency-independent phase shifters (PSs). For simplification, we define $\mathcal{D}=\{1, \dots, D\}$ as the index set of the TD elements. Therefore, the transmitted signal $\mathbf{x}_{a,m}$ by AP $a$ at SC $m$ is denoted as
\begin{equation}\label{equ:x}
	\mathbf{x}_{a,m}=\mathbf{F}_{\mathrm{RF},a}\mathbf{F}_{\mathrm{TD},a,m}\sum_{j\in \mathcal{K}}{\mathbf{w}_{a,j,m}s_{j,m}},
\end{equation}
where $\mathbf{F}_{\mathrm{RF},a} \triangleq \left[ \mathbf{F}_{\mathrm{RF},a,1},\dots ,\mathbf{F}_{\mathrm{RF},a,N_{\mathrm{RF}}} \right] \in \mathbb{C} ^{N_{\mathrm{Tx}}\times DN_{\mathrm{RF}}}$ denotes the analog beamformer, where
\begin{equation}
\begin{split}
\mathbf{F}_{\mathrm{RF},a,r}&\triangleq \mathrm{col}\mathrm{diag}\left\{ \left[ \mathbf{f}_{\mathrm{RF},a,r,1},\dots ,\mathbf{f}_{\mathrm{RF},a,r,D} \right] \right\} \in \!\mathbb{C} ^{N_{\mathrm{Tx}}\!\times \!D}
\\
&=\left[ \begin{matrix}
	\mathbf{f}_{\mathrm{RF},a,r,1}&		\mathbf{0}_{N_{\mathrm{D}}}&		\cdots&		\mathbf{0}_{N_{\mathrm{D}}}\\
	\mathbf{0}_{N_{\mathrm{D}}}&		\mathbf{f}_{\mathrm{RF},a,r,2}&		\cdots&		\mathbf{0}_{N_{\mathrm{D}}}\\
	\vdots&		\vdots&		\ddots&		\vdots\\
	\mathbf{0}_{N_{\mathrm{D}}}&		\mathbf{0}_{N_{\mathrm{D}}}&		\cdots&		\mathbf{f}_{\mathrm{RF},a,r,D}\\
\end{matrix} \right] 
\end{split}
\end{equation}
consists of $D$ analog beamformer vectors $\mathbf{f}_{\mathrm{RF},a,r,d} \in\mathbb{C} ^{N_{\mathrm{D}}}$, which satisfies the modulus constraint $\left| \mathbf{f}_{\mathrm{RF},a,r,d} \right|=\frac{1}{\sqrt{N_{\mathrm{D}}}}$.
Besides, $\mathbf{F}_{\mathrm{TD},a,m} \triangleq \mathrm{col}\mathrm{diag}\left\{ \left[ \mathbf{f}_{\mathrm{TD},a,1,m},\cdots ,\mathbf{f}_{\mathrm{TD},a,N_{\mathrm{RF}},m} \right] \right\} \in \mathbb{C} ^{DN_{\mathrm{RF}}\times N_{\mathrm{RF}}}$ is the TD matrix, which consists of $N_{\mathrm{RF}}$ TD vectors $\mathbf{f}_{\mathrm{TD},a,r,m}\triangleq e^{-j2\pi f_m\mathbf{t}_{a,r}^{\mathrm{AP}}}\in \mathbb{C} ^D$ where $\mathbf{t}_{a,r}^{\mathrm{AP}}\triangleq \left[ t_{a,r,1}^{\mathrm{AP}},\cdots ,t_{a,r,D}^{\mathrm{AP}} \right] \in \mathbb{C} ^D$ represents the TD factor vector. Additionally, $s_{j,m}\in \mathbb{C}$ and $\mathbf{w}_{a,j,m} \in \mathbb{C}^{N_{\mathrm{RF}}}$ denote the pilot symbol and baseband beamformer, respectively. Notice that the pilot signal $\mathbf{s}_m \triangleq \left[s_{1,m}, \cdots, s_{K,m}\right]^\mathsf{T} \in \mathbb{C}^{K}$ satisfies the constraint $\mathbb{E}\left\{\mathbf{s}_m \mathbf{s}_m^\mathsf{H}\right\}=\mathbf{I}_K$. Let $P_{a,\max}$ denote the maximum transmit power, the power constraint for AP $a$ can be given by $\sum_{j\in \mathcal{K}}{\sum_{m\in \mathcal{M}}{\left\| \mathbf{F}_{\mathrm{RF},a}\mathbf{F}_{\mathrm{TD},a,m}\mathbf{w}_{a,m,j} \right\| ^2}}\le P_{a,\max}$.

\subsection{System Model}
The operation channel $\mathbf{h}_{a,k,m}^{\mathsf{H}}$ between AP $a$ and UE $k$ at SC $m$ is expressed as 
\begin{equation}\label{equ:h}
\begin{split}
	\mathbf{h}_{a,k,m}^{\mathsf{H}}&=\mathbf{h}_{\mathrm{dir},a,k,m}^{\mathsf{H}}+\sum_{r\in \mathcal{R}}{\mathbf{u}_{r,k,m}^{\mathsf{H}}\mathbf{\Theta }_r\mathbf{T}_{r,m}\mathbf{G}_{a,r,m}}
	\\
	&=\underset{\mathrm{AP}\text{-}\mathrm{UE}\,\mathrm{link}}{\underbrace{\mathbf{h}_{\mathrm{dir},a,k,m}^{\mathsf{H}}}}+\underset{\mathrm{AP}\text{-}\mathrm{RIS}\text{-}\mathrm{UE}\,\mathrm{links}}{\underbrace{\mathbf{u}_{k,m}^{\mathsf{H}}\mathbf{\Theta T}_m\mathbf{G}_{a,m}}},
\end{split}
\end{equation}
where $\mathbf{h}_{\mathrm{dir},a,k,m}^{\mathsf{H}}\in\mathbb{C}^{N_{\mathrm{Tx}}}$, $\mathbf{u}_{r,k,m}^{\mathsf{H}}\in\mathbb{C}^{N_{\mathrm{RIS}}}$ and $\mathbf{G}_{a,r,m}\in\mathbb{C}^{N_{\mathrm{RIS}} \times N_{\mathrm{Tx}}}$
denote the direct channel from AP $a$ to UE $k$, sub-channel from RIS $r$ to UE $k$ and sub-channel from AP $a$ to RIS $r$, respectively. $\mathbf{\Theta }_m\triangleq \mathrm{diag}\left\{ \theta _{r,1},...,\theta _{r,N_{\mathrm{RIS}}} \right\} \in \mathbb{C} ^{N_{\mathrm{RIS}}\times N_{\mathrm{RIS}}}$ and $\mathbf{T}_{r,m}\triangleq \mathrm{diag}\left\{ t_{r,m,1}^{\mathrm{RIS}},...,t_{r,m,N_{\mathrm{RIS}}}^{\mathrm{RIS}} \right\} \in \mathbb{C} ^{N_{\mathrm{RIS}}\times N_{\mathrm{RIS}}}$ denote the phase shift matrix and time delay matrix of RIS $r$, respectively. In practical, the low-resolution phase of $\theta _{r,n}$ is discrete, satisfying $\theta _{r,n}\in \mathcal{F}$ with 
\begin{equation}
\mathcal{F} \triangleq \left\{ \theta _{r,n}\left| \theta _{r,n}\in \left\{ 1,e^{j\frac{2\pi}{F}},\cdots ,e^{j\frac{2\pi (F-1)}{F}} \right\} \right. \right\} ,
\end{equation}
where $F$ is the discrete level \cite{di2020hybrid}. Besides, $\mathbf{\Theta }\triangleq \mathrm{diag}\left\{ \mathbf{\Theta }_1,...,\mathbf{\Theta }_R \right\} $ and $\mathbf{T}_{m}\triangleq\mathrm{diag}\left\{ \mathbf{T}_{1,m},...,\mathbf{T}_{R,m} \right\}$.

Therefore, the received signal $y_{k,m}$ by UE $k$ at SC $m$ can be written as
\begin{equation}\label{equ:y}
	y_{k,m}=\sum_{a\in \mathcal{A}}{\mathbf{h}_{a,k,m}^{\mathsf{H}}\bar{\mathbf{F}}_{a,m}\mathbf{x}_{a,m}}+n_{k,m},
\end{equation}
where $\bar{\mathbf{F}}_{a,m}\triangleq \mathbf{F}_{\mathrm{RF},a}\mathbf{F}_{\mathrm{T},a,m}\in \mathbb{C} ^{N_{\mathrm{Tx}}\times N_{\mathrm{RF}}}$ denotes the frequency-dependent analog beamformer, $n_{k,m} \sim \mathcal{C N}\left(0, \sigma_{k,m}^2\right)$ is the additive white Gaussian noise (AWGN). 
Furthermore, by substituting \eqref{equ:x} into \eqref{equ:y}, the received signal $y_{k,m}$ can be partitioned into two items, given by
\begin{equation}\label{equ:y_2part}
	y_{k,m}=\underset{\mathrm{Signal}}{\underbrace{\mathbf{h}_{k,m}^{\mathsf{H}}\bar{\mathbf{F}}_m\mathbf{w}_{k,m}s_{k,m}}}+\underset{\mathrm{Interfrence}\&\mathrm{Noise}}{\underbrace{\sum_{j\in \mathcal{K} \backslash k}{\mathbf{h}_{k,m}^{\mathsf{H}}\bar{\mathbf{F}}_m\mathbf{w}_{j,m}s_{j,m}+n_{k,m}}}},
\end{equation}
where $\mathbf{h}_{k,m}\triangleq[ \mathbf{h}_{1,k,m}^{\mathsf{T}},\dots ,\mathbf{h}_{A,k,m}^{\mathsf{T}} ] ^{\mathsf{T}}\in \mathbb{C} ^{AN_{\mathrm{Tx}}}
$, $\mathbf{w}_{k,m}\triangleq[ \mathbf{w}_{1,k,m}^{\mathsf{T}},\dots ,\mathbf{w}_{A,k,m}^{\mathsf{T}} ] ^{\mathsf{T}}\in \mathbb{C} ^{AN_{\mathrm{RF}}}$ and  
\begin{equation}
\begin{split}
	\bar{\mathbf{F}}_m&\triangleq \mathrm{blk}\mathrm{diag}\left\{ \left[ \bar{\mathbf{F}}_{1,m},\dots ,\bar{\mathbf{F}}_{A,m} \right] \right\}  \in \mathbb{C} ^{AN_{\mathrm{Tx}}\times AN_{\mathrm{RF}}}
	\\
	&=\left[ \begin{matrix}
		\bar{\mathbf{F}}_{1,m}&		\mathbf{0}_{N_{\mathrm{Tx}}\times N_{\mathrm{RF}}}&		\cdots&		\mathbf{0}_{N_{\mathrm{Tx}}\times N_{\mathrm{RF}}}\\
		\mathbf{0}_{N_{\mathrm{Tx}}\times N_{\mathrm{RF}}}&		\bar{\mathbf{F}}_{2,m}&		\cdots&		\mathbf{0}_{N_{\mathrm{Tx}}\times N_{\mathrm{RF}}}\\
		\vdots&		\vdots&		\ddots&		\vdots\\
		\mathbf{0}_{N_{\mathrm{Tx}}\times N_{\mathrm{RF}}}&		\mathbf{0}_{N_{\mathrm{Tx}}\times N}&		\cdots&		\bar{\mathbf{F}}_{A,m}\\
	\end{matrix} \right].
\end{split}
\end{equation}
It is noticed that the former item in \eqref{equ:y_2part} denotes the desired signal for UE $k$ while the latter item represents the interference from other UEs and the noise.

\subsection{Channel Model}
In THz band, since the path gains of non-line-of-sight (NLoS) paths are much lower than that of LoS path, we only consider LoS channel model \cite{he2024wireless}. Accordingly, the channel $\mathbf{G}_{a,r,m}$ from AP $a$ to RIS $r$ at SC $m$ is expressed as
\begin{equation}\label{equ:G}
\mathbf{G}_{a,r,m}=\alpha _{a,r}e^{-j2\pi \tau _{a,r}^{\mathrm{G}}f_m}\mathbf{b}_{\mathrm{RIS},m}\left( \vartheta _{a,r},\varphi _{a,r} \right) \mathbf{a}_{\mathrm{Tx},m}^{\mathsf{H}}\left( \phi _{a,r} \right), 
\end{equation}
where $\alpha _{a,r}$ and $\tau _{a,r}^{\mathrm{G}}$ represent the complex gain and  time delay. $\vartheta _{a,r}$, $\varphi _{a,r}$ and $\phi _{a,r}$ are the azimuth of arrival (AoA) at RIS, elevation of arrival (EoA) at RIS, and azimuth of departure (AoD) at AP, respectively. Besides, $f_m$ is the frequency of SC $m$, given by
\begin{equation}\label{equ:f_m}
	f_m=f_c+\frac{B}{M}\left(m-1-\frac{M-1}{2}\right), \forall m=1,2, \ldots, M,
\end{equation}
where $f_c$ is the central frequency. Additionally, $\mathbf{a}_{\mathrm{Tx},m}\in \mathbb{C} ^{N_{\mathrm{Tx}}} $ and $\mathbf{b}_{\mathrm{RIS},m}\in \mathbb{C} ^{N_{\mathrm{RIS}}}$ are the beam-split-affected array response vectors (ARVs) of ULA and UPA, which can be written as
\begin{subequations}\label{equ:ARV}
	\begin{equation}
		\mathbf{a}_{\mathrm{Tx},m}\left( \phi \right) =\frac{1}{\sqrt{N_{\mathrm{Tx}}}}e^{-j\pi \eta _m\mathbf{n}_t\sin \phi},
	\end{equation}
	\begin{equation}
		\mathbf{b}_{\mathrm{RIS},m}\!\left( \vartheta ,\varphi \right) \!=\!\!\frac{1}{\sqrt{N_{\mathrm{RIS}}}}e^{\!-j\pi \eta _m\!\mathbf{n}_x\!\sin\! \vartheta\! \cos\!\varphi}\otimes e^{\!-j\pi \eta _m\!\mathbf{n}_y\!\sin\! \vartheta \!\sin\! \varphi},
	\end{equation}
\end{subequations}
where $\mathbf{n}_t=\left[ 0,\cdots ,N_{\mathrm{Tx}}-1 \right] $, $\mathbf{n}_x=\left[ 0,\cdots ,N_{\mathrm{X}}-1 \right] $ and $\mathbf{n}_y=\left[ 0,\cdots ,N_{\mathrm{Y}}-1 \right] $. Similarly, the channel $\mathbf{u}_{r,k,m}$ from RIS $r$ to UE $k$ at SC $m$ is expressed as 
\begin{equation}\label{equ:u}
	\mathbf{u}_{r,k,m}=\beta _{r,k}e^{-\mathrm{j}2\pi \tau _{r,k}^{\mathrm{u}}f_m}\mathbf{b}_{\mathrm{RIS},m}\left( \mu _{r,k},\nu _{r,k} \right) ,
\end{equation}
where $\beta _{r,k}$, $\tau _{r,k}^{\mathrm{u}}$, $\mu _{r,k}$ and $\nu _{r,k}$ represent the complex gain, time delay, AoD and elevation of departure (EoD) at RIS, respectively. Similarly, the direct channel $\mathbf{h}_{\mathrm{dir},a,k,m}$ from AP $a$ to UE $k$ at SC $m$ is expressed as 
\begin{equation}\label{equ:h_dir}
	\mathbf{h}_{\mathrm{dir},a,k,m}=\gamma _{a,k}e^{-j2\pi \tau _{a,k}^{\mathrm{d}}f_m}\mathbf{a}_{\mathrm{Tx},m}\left( \psi _{a,k} \right),
\end{equation}
where $\gamma _{a,k}$, $\tau _{a,k}^{\mathrm{d}}$ and $\psi _{a,k}$ represent the complex gain, time delay and AoD at AP, respectively. 

\subsection{Problem Formulation}\label{sec:Prob_Form}
To maximize the WSR of RIS-assisted THz cell-free mMIMO system, the signal-to-interference-plus-noise ratio (SINR) for each UE is firstly calculated. Based on \eqref{equ:y_2part}, the SINR for UE $k$ at SC $m$ is denoted as
\begin{equation}\label{equ:SINR}
	\mathsf{SINR}_{k,m}=\frac{\left| \mathbf{h}_{k,m}^{\mathsf{H}}\bar{\mathbf{F}}_m\mathbf{w}_{k,m} \right|^2}{\sum_{j\in \mathcal{K} \backslash k}{\left| \mathbf{h}_{k,m}^{\mathsf{H}}\bar{\mathbf{F}}_m\mathbf{w}_{j,m} \right|^2}+\sigma _{k,m}^{2}}.
\end{equation}
Subsequently, accumulating the SINR of all $K$ UEs at all $M$ SCs, the WSR can be expressed as 
\begin{equation}\label{equ:WSR}
	R_{\mathrm{sum}}=\sum_{k\in \mathcal{K}}{\sum_{m\in \mathcal{M}}{\varrho _k\log _2\left( 1+\mathsf{SINR}_{k,m} \right)}},
\end{equation}
where $\varrho _k \in \mathbb{R}^{+}$ represents the weight of UE $k$. Given the WSR in \eqref{equ:WSR}, the precoding problem is thus formulated as a WSR maximization optimization problem, given by
\begin{equation}\label{equ:P1}
\begin{split}
		\left( \mathrm{P}1 \right) \,\,&\max_{\bar{\mathbf{F}}_m,\mathbf{w},\mathbf{\Theta },\mathbf{T }_m}
		 f_1\left( \bar{\mathbf{F}}_m,\mathbf{w},\mathbf{\Theta },\mathbf{T }_m \right) =R_{\mathrm{sum}}
		\\
		&\mathrm{s}.\mathrm{t}. \,\,C_1:\sum_{k\in \mathcal{K}}{\sum_{m\in \mathcal{M}}{\left\| \bar{\mathbf{F}}_{a,m}\mathbf{w}_{a,k,m} \right\| ^2}}\le P_{a,\max},\forall a\in \mathcal{A} 
		\\
		&\,\,\,\,\,\,\,\,\,\,C_2:\theta _{r,n}\in \mathcal{F} ,\forall r\in \mathcal{R} ,n=1,\dots ,N_{\mathrm{RIS}}
		\\
		&\,\,\,\,\,\,\,\,\,\,C_3:\left| \mathbf{f}_{\mathrm{RF},a,r,d} \right|=\frac{1}{\sqrt{N_{\mathrm{D}}}},\forall a\in \mathcal{A} ,r\in \mathcal{R} ,d\in \mathcal{D} \\
		&\,\,\,\,\,\,\,\,\,\,C_4:\left| t_{r,m,n}^{\mathrm{RIS}} \right|=1 ,\forall r\in \mathcal{R},m\in \mathcal{M},n=1,...,\!N_{\mathrm{RIS}},
\end{split}
\end{equation}
where $\mathbf{w}\triangleq\left[ \mathbf{w}_{1,1}^{\mathsf{T}},\dots ,\mathbf{w}_{K,1}^{\mathsf{T}},\dots ,\mathbf{w}_{K,M}^{\mathsf{T}} \right] ^{\mathsf{T}}\in \mathbb{C} ^{AKMN_{\mathrm{RF}}}$.
\section{Proposed Optimization Framework}\label{sec:Proposed Method}
In this section, we develop an alternate optimization (AO) framework to solve the initial optimization problem formulated in $\left( \mathrm{P}1 \right)$, as presented in Algorithm \ref{Alg:AO}. Based on fact that the location of RISs remain unchanged, we first adjust the TDs at APs to align the analog beamformer with RISs. Then, we introduce auxiliary variables $\boldsymbol{\rho}$ to decompose the initial complex optimization problem into separated sub-problems. Finally, we alternatively optimize the delay phase hybrid precoding architecture $\bar{\mathbf{F}}_m$ and $\mathbf{w}$ of AP, phase shift matrix $\mathbf{\Theta }$ and
time delay matrix $\mathbf{T}_m$ of RIS to solve the remaining sub-problems.

\begin{algorithm}[t]
	\caption{The Proposed Optimization Framework}
	\label{Alg:AO}
	\begin{algorithmic}[1]
			\STATE\textbf{Initialize}: $\bar{\mathbf{F}}_m,\mathbf{w},\mathbf{\Theta },\mathbf{T}_m$.
			\STATE Update frequency-dependent beamformer matrix $\bar{\mathbf{F}}_m$ by \eqref{equ:F*_TD} and \eqref{equ:F_a,rOpt}.
			\WHILE {$n \leq N_\mathrm{iter}$}
				\STATE Update auxiliary variables $\boldsymbol{\rho}$ by \eqref{equ:rho_*}.
				\STATE Update auxiliary variables $\boldsymbol{\lambda}$ by \eqref{equ:lambda_*}.
				\STATE Update baseband beamformer matrix $\mathbf{w}$ by \eqref{equ:P4}.
				\STATE Update auxiliary variables $\boldsymbol{\omega}$ by \eqref{equ:omega_*}.
				\STATE Update RIS phase shift matrix $\mathbf{\Theta }$ by \eqref{equ:P6}.
				\STATE Update auxiliary variables $\boldsymbol{\gamma}$ by \eqref{equ:gamma_*}.
				\STATE Update TD matrix of RIS $\mathbf{T}_m$ by \eqref{equ:P8}.
				\IF {$R_{\mathrm{sum}}$ converges}
					\STATE break.
				\ENDIF
			\ENDWHILE
		\end{algorithmic}
\end{algorithm}

\subsection{Precoding Design for Frequency-dependent Analog Beamformer $\bar{\mathbf{F}}_m$}
To compensate the severe array gain loss, the beams generated by the analog beamformers should be aligned with the target physical directions \cite{dai2022delay}. Therefore, the beam generated by the $r$-th frequency-dependent analog beamformer vector $\bar{\mathbf{f}}_{a,r,m}
=\mathbf{F}_{\mathrm{RF},a,r}\mathbf{f}_{\mathrm{TD},a,r,m}\in \mathbb{C} ^{N_{\mathrm{Tx}}}$ is aligned with RIS $r$, given by 
\begin{equation}\label{equ:barF_r}
\bar{\mathbf{f}}_{a,r,m}=\left[ \bar{\mathbf{f}}_{a,r,1}^{\mathsf{T}},\dots ,\bar{\mathbf{f}}_{a,r,D}^{\mathsf{T}} \right] ^{\mathsf{T}}=\mathbf{a}_{\mathrm{Tx},c}\left( \phi _{a,r} \right) . 
\end{equation}
According to \eqref{equ:barF_r}, the TD factor vector of $\mathbf{f}_{\mathrm{TD},a,r,m}$ is set to align $\bar{\mathbf{f}}_{a,r,m}$ with $\phi _{a,r}$ across all $M$ SCs, which can be calculated as
\begin{equation}\label{equ:t_AP}
\mathbf{t}_{a,r}^{\mathrm{AP}}=\left( \eta _m-1 \right) N_{\mathrm{D}}\sin\phi _{a,r}\mathbf{d}/f_m ,
\end{equation}
where $\mathbf{d}=\left[ 0,\cdots ,D-1 \right]$. However, due to the hardware constraint, the TD factor vector satisfies $\mathbf{t}_{a,r}^{\mathrm{AP}}=\Delta _{a,r}T_c\mathbf{d}$, where $T_c=1/f_c$ denotes the period of the central carrier and $\Delta _{a,r}$ is the number of period that should be delayed for the path component between AP $a$ and RIS $r$. Therefore, according to \eqref{equ:t_AP}, $\Delta _{a,r}$ is easily obtained as 
\begin{equation}\label{equ:Delta_a,r}
\Delta _{a,r}=\frac{\left( \eta _m-1 \right) \sin\phi _{a,r}N_{\mathrm{D}}}{2\eta _m}.
\end{equation}
According to \eqref{equ:Delta_a,r}, the phase shift of $\mathbf{f}_{\mathrm{TD},a,r,m}$ can be divided into two parts $-\pi\eta _m\sin\phi _{a,r}N_{\mathrm{D}}$ and $\pi\sin\phi _{a,r}N_{\mathrm{D}}$. It is obvious that the former part is frequency-dependent and can be realized by setting $\Delta _{a,r}=\sin\phi _{a,r}N_{\mathrm{D}}/2$. Consequently, the optimal TD vector can be expressed as
\begin{equation}\label{equ:F*_TD}
\mathbf{f}_{\mathrm{TD},a,r,m}^{\star}=e^{-j\pi \eta _mN_{\mathrm{D}}\sin \phi _{a,r}\mathbf{d}}.
\end{equation}
In contrast, the latter part is frequency-independent, which can be achieved independently without TDs. In this case, an extra phase shift is added on the analog beamformer $\mathbf{F}_{\mathrm{RF},a,r}$ as compensation, given by 
\begin{equation}\label{equ:F_a,rOpt}
\mathbf{F}^{\star}_{\!\mathrm{RF\!},a,r}\!=\mathrm{coldiag}\!\left\{ \!\left[ \mathbf{f}_{\mathrm{RF\!},a,r,1}\!,\cdots\!,\!\mathbf{f}_{\mathrm{RF\!},a,r,D}e^{j\!\pi \!\left( D-1 \right)\! \sin \phi _{a,r}\!N_{\mathrm{D}}} \right] \!\right\}.
\end{equation}
So far, the optimal frequency-dependent analog beamformer vector can be easily calculated by $\bar{\mathbf{f}}^{\star}_{a,r,m}
=\mathbf{F}^{\star}_{\mathrm{RF},a,r}\mathbf{f}^{\star}_{\mathrm{TD},a,r,m}$.

\subsection{Optimization Problem Decoupling}
The non-convex optimization problem formulated in $\left( \mathrm{P}1 \right)$ can be decoupled by utilizing Lagrangian dual reformulation (LDR). Specifically, by substituting the optimal frequency-dependent analog beamformer $\bar{\mathbf{F}}^{\star}_{m}$ and introducing auxiliary variables $\boldsymbol{\rho }\triangleq\left[ \rho _{1,1},\rho _{1,2},\dots ,\rho _{1,K},\dots ,\rho _{M,K} \right] ^{\mathsf{T}}$, $\left( \mathrm{P}1 \right)$ is equivalent to
\begin{equation}\label{equ:P2}
\begin{split}
\left( \mathrm{P}2 \right) \,\,&\max_{\mathbf{w},\mathbf{\Theta },\mathbf{T}_m,\boldsymbol{\rho }} \,\,f_2\left( \mathbf{w},\mathbf{\Theta },\mathbf{T}_m,\boldsymbol{\rho } \right) 
\\
&\mathrm{s}.\mathrm{t}.\,\, C_1:\sum_{k\in \mathcal{K}}{\sum_{m\in \mathcal{M}}{\left\| \bar{\mathbf{F}}_{a,m}^{\star}\mathbf{w}_{a,k,m} \right\| ^2}}\le P_{a,\max},\forall a\in \mathcal{A} 
\\
&\,\,\,\,\,\,\,\,\,\,C_2:\theta _{r,n}\in \mathcal{F} ,\forall r\in \mathcal{R} ,n=1,\dots ,N_{\mathrm{RIS}}\\
&\,\,\,\,\,\,\,\,\,\,C_3:\left| t_{r,m,n}^{\mathrm{RIS}} \right|=1 ,\forall r\in \mathcal{R},m\in \mathcal{M},n=1,...,\!N_{\mathrm{RIS}},
\end{split}
\end{equation}
where
\begin{equation}\label{equ:f2}
\begin{split}
f_2\!\left( \mathbf{w},\mathbf{\Theta },\mathbf{T}_m,\boldsymbol{\rho } \right) \!&=\!\!\sum_{k\in \mathcal{K}}\!{\sum_{m\in \mathcal{M}}{\!\!\!\varrho_k\ln \!\left( \!1\!+\!\rho _{k,m} \right)}}\!-\!\!\sum_{k\in \mathcal{K}}\!{\sum_{m\in \mathcal{M}}{\!\varrho_k\rho _{k,m}}}
\\
&+\sum_{k\in \mathcal{K}}{\sum_{m\in \mathcal{M}}{\varrho_k\left( 1+\rho _{k,m} \right) g_2\left( \mathbf{w},\mathbf{\Theta },\mathbf{T}_m \right)}},
\end{split}
\end{equation}
where 
\begin{equation}\label{equ:g2}
g_2\left(\mathbf{w}, \mathbf{\Theta } \right) =\frac{\left| \mathbf{h}_{k,m}^{\mathsf{H}}\bar{\mathbf{F}}_{m}^{\star}\mathbf{w}_{k,m} \right|^2}{\sum_{j\in \mathcal{K}}{\left| \mathbf{h}_{k,m}^{\mathsf{H}}\bar{\mathbf{F}}_{m}^{\star}\mathbf{w}_{j,m} \right|^2}+\sigma _{k,m}^{2}}.
\end{equation}
Fixing $\left( \mathbf{w},\mathbf{\Theta },{\mathbf{T}_{m}} \right) $, the optimal $\boldsymbol{\rho}^{\star}$ in \eqref{equ:f2} is at the point of zero derivative $\partial f_2\left( \boldsymbol{\rho } \right) /\partial \rho _{k,m}=0$. Therefore, the solution can be easily calculated as 
\begin{equation}\label{equ:rho_*}
\begin{split}
	\rho _{k,m}^{\star}&=\frac{\left| \mathbf{h}_{k,m}^{\mathsf{H}}\bar{\mathbf{F}}_{m}^{\star}\mathbf{w}_{k,m} \right|^2}{\sum_{j\in \mathcal{K} \backslash k }{\left| \mathbf{h}_{k,m}^{\mathsf{H}}\bar{\mathbf{F}}_{m}^{\star}\mathbf{w}_{j,m} \right|^2}+\sigma _{k,m}^{2}}
	\\
	&=\mathsf{SINR}_{k,m}\left(\bar{\mathbf{F}}_{m}^{\star}\right).
\end{split}
\end{equation}
So far, given the optimal $\boldsymbol{\rho}^{\star}$, the optimization of $\left( \mathbf{w},\mathbf{\Theta },{\mathbf{T}_{m}} \right)$ is only relevant to the final part in \eqref{equ:P2}. 

\subsection{Precoding Design for Baseband Beamformer $\mathbf{w}$}
Fixing $\left( \bar{\mathbf{F}}_m^{\star},\mathbf{\Theta },\mathbf{T}_{m},\boldsymbol{\rho } \right) $, it is observed that the optimization of $\mathbf{w}$ is only relevant to the third part in \eqref{equ:f2}. Furthermore, we exploit the multidimensional complex quadratic transform (MCQT) \cite{shen2018fractional}, which introduces auxiliary variables $\boldsymbol{\lambda }\triangleq\left[ \lambda _{1,1},\cdots ,\lambda _{1,K},\cdots ,\lambda _{M,K} \right]$.
Therefore, the optimization subproblem of $\mathbf{w}$ can be formulated as
\begin{equation}\label{equ:P3}
\begin{split}
\left( \mathrm{P}3 \right) \,\,&\max_{\mathbf{w},\boldsymbol{\lambda }} \,\,f_3\left( \mathbf{w},\boldsymbol{\lambda } \right) 
\\
\,\,&\mathrm{s}.\mathrm{t}.\sum_{k\in \mathcal{K}}{\sum_{m\in \mathcal{M}}{\left\| \bar{\mathbf{F}}_{a,m}^{\star}\mathbf{w}_{a,k,m} \right\| ^2}}\le P_{a,\max},\forall a\in \mathcal{A} ,
\end{split}
\end{equation}
where
\begin{equation}\label{equ:f3}
\begin{split}
f_3\left( \mathbf{w},\boldsymbol{\lambda } \right) =\sum_{k\in \mathcal{K}}{2\varsigma _{k,m}\mathfrak{R} \mathfrak{e} \left\{ \lambda _{k,m}^{*}\mathbf{h}_{k,m}^{\mathsf{H}}\bar{\mathbf{F}}_{m}^{\star}\mathbf{w}_{k,m} \right\}}
\\-\sum_{k\in \mathcal{K}}{\sum_{m\in \mathcal{M}}{\left| \lambda _{k,m} \right|^2\left( \sum_{j\in \mathcal{K}}{\left| \mathbf{h}_{k,m}^{\mathsf{H}}\bar{\mathbf{F}}_{m}^{\star}\mathbf{w}_{j,m} \right|^2}+\sigma _{k,m}^{2} \right)}},
\end{split}
\end{equation}
where $\varsigma _{k,m}\triangleq\sqrt{\varrho _k\left( 1+\rho _{k,m} \right)}$. To solve subproblem $\left( \mathrm{P}3 \right)$, we update $ \mathbf{w}$ and $\boldsymbol{\lambda }$ alternatively. 

\subsubsection{Fix $\mathbf{w}$ and Solve $\boldsymbol{\lambda }^{\star}$}
Firstly, with $ \mathbf{w}$ fixed, the optimal $\boldsymbol{\lambda }^{\star}$ is at the point of zero derivative $\frac{\partial f_3\left( \boldsymbol{\lambda } \right)}{\partial \lambda _{k,m}}=0
$, which is calculated as
\begin{equation}\label{equ:lambda_*}
\lambda _{k,m}^{\star}=\frac{\varsigma _{k,m}}{1+1/\mathsf{SINR}_{k,m}\left(\bar{\mathbf{F}}_{m}^{\star}\right)}
\end{equation}

\subsubsection{Fix $\boldsymbol{\lambda }$ and Solve $\mathbf{w}^{\star}$}
Then, by substituting the fixed $\boldsymbol{\lambda }$ into \eqref{equ:f3}, the objective function can be transformed into a matrix form representation of $\mathbf{w}$, which is expressed as
\begin{equation}\label{equ:f3_matrix}
f_3\left( \mathbf{w} \right)=-\mathbf{w}^{\mathsf{H}}\mathbf{\Xi w}+2\mathfrak{R} \mathfrak{e} \left\{ \tilde{\boldsymbol{\xi}}^{\mathsf{H}}\mathbf{w} \right\} -\zeta 
\end{equation}
where $\mathbf{\Xi }\triangleq\mathrm{diag}\left\{ \mathbf{I}_K\otimes \mathbf{\Xi }_1,\dots ,\mathbf{I}_K\otimes \mathbf{\Xi }_M \right\}$, $\mathbf{\Xi }_m\triangleq\sum_{k\in \mathcal{K}}{\boldsymbol{\xi }_{k,m}\boldsymbol{\xi }_{k,m}^{\mathsf{H}}}$ and $\boldsymbol{\xi }_{k,m}\triangleq\lambda _{k,m}\left( \bar{\mathbf{F}}_{m}^{\star}\right) ^{\mathsf{H}}\mathbf{h}_{k,m}$. Besides, $\tilde{\boldsymbol{\xi}}\triangleq[ \varsigma _{1,1}\boldsymbol{\xi }_{1,1}^{\mathsf{T}},\dots ,\varsigma _{K,1}\boldsymbol{\xi }_{K,1}^{\mathsf{T}},\dots ,\varsigma _{K,M}\boldsymbol{\xi }_{K,M}^{\mathsf{T}}] $ and $\zeta \triangleq\sum_{k\in \mathcal{K}}{\sum_{m\in \mathcal{M}}{\left| \lambda _{k,m} \right|^2\sigma _{k,m}^{2}}}$. Therefore, the optimization of $\mathbf{w}$ can be formulated as 
\begin{equation}\label{equ:P4}
\begin{split}
\left( \mathrm{P}4 \right) \,\,&\min_{\mathbf{w}} \,\,f_4\left( \mathbf{w} \right) =\mathbf{w}^{\mathsf{H}}\mathbf{\Xi w}-2\mathfrak{R} \mathfrak{e} \left\{ \tilde{\boldsymbol{\xi}}^{\mathsf{H}}\mathbf{w} \right\} 
\\
&\mathrm{s}.\mathrm{t}.\mathbf{w}^{\mathsf{H}}\mathbf{\Upsilon }_a\mathbf{w}\le P_{a,\max},\forall a\in \mathcal{A} ,
\end{split}
\end{equation}
where $\mathbf{\Upsilon }_a=\mathrm{blk}\mathrm{diag}\left\{ \mathbf{\Upsilon }_{a,1},\dots ,\mathbf{\Upsilon }_{a,M} \right\} $ and $\mathbf{\Upsilon }_{a,m}=\mathbf{I}_K\otimes [ \left( \mathbf{e}_a\mathbf{e}_{a}^{\mathsf{H}} \right) \otimes \left( \left( \bar{\mathbf{F}}_{a,m}^{\star} \right) ^{\mathsf{H}}\bar{\mathbf{F}}_{a,m}^{\star} \right) ] $. Since the matrices $\mathbf{\Xi}$ and $\left\{ \mathbf{\Upsilon }_a \right\} _{a=1}^{A}$ are all positive semidefinite, $\left( \mathrm{P}4 \right) $ is a convex optimization problem and can be easily solved by existing solutions such as primal-dual subgradient (PDS) \cite{zhang2023joint}.

\subsection{Precoding Design for RIS Phase Shift $\mathbf{\Theta}$}
Fixing $\left(\bar{\mathbf{F}}_m^{\star}, \mathbf{w },\mathbf{T}_{m},\boldsymbol{\rho }\right) $, the optimization of $\mathbf{\Theta}$ is only relevant to the third part in \eqref{equ:f2}. Similar to the optimization of $\mathbf{w }$, we exploit the MCQT again. By introducing auxiliary variables $\boldsymbol{\omega }\triangleq\left[ \omega_{1,1},\cdots ,\omega _{K,1},\cdots ,\omega _{K,M} \right]^{\mathsf{T}}$, the optimization subproblem of $\mathbf{\Theta}$ can be formulated as
\begin{equation}\label{equ:P5}
\begin{split}
	\left( \mathrm{P}5 \right) \,\,&\max_{\mathbf{\Theta },\boldsymbol{\omega } } \,\,f_5\left( \mathbf{\Theta },\boldsymbol{\omega }  \right) =\sum_{k\in \mathcal{K}}{\sum_{m\in \mathcal{M}}{g_5\left( \mathbf{\Theta },\boldsymbol{\omega } \right)}}
	\\
	& \mathrm{s}.\mathrm{t}. \left| \theta _{r,n} \right|\in \mathcal{F} , \forall r\in \mathcal{R} , n=1,\dots ,N_{\mathrm{RIS}}	,
\end{split}
\end{equation}
where 
\begin{equation}\label{equ:g5}
\begin{split}
	g_5\left( \mathbf{\Theta },\boldsymbol{\omega } \right) =2\varsigma _{k,m}\mathfrak{R} \mathfrak{e} \left\{ \omega _{k,m}^{*}\mathbf{h}_{k,m}^{\mathsf{H}}\bar{\mathbf{F}}_{m}^{\star}\mathbf{w}_{k,m}\right\} \\
	-\left| \omega _{k,m} \right|^2\left( \sum_{j\in \mathcal{K}}{\left| \mathbf{h}_{k,m}^{\mathsf{H}}\bar{\mathbf{F}}_{m}^{\star}\mathbf{w}_{j,m} \right|^2}+\sigma _{k,m}^{2} \right) .
\end{split}
\end{equation}
To solve subproblem $\left( \mathrm{P}5 \right)$, we update $ \mathbf{\Theta}$ and $\boldsymbol{\omega }$ alternatively. 

\subsubsection{Fix $ \mathbf{\Theta}$ and Solve $\boldsymbol{\omega }^{\star}$}
Firstly, with $ \mathbf{\Theta}$ fixed, the optimal $\boldsymbol{\omega }^{\star}$ is at the point of zero derivative $\frac{\partial f_5\left( \boldsymbol{\omega } \right)}{\partial \omega _{k,m}}=0
$, which is calculated as
\begin{equation}\label{equ:omega_*}
\omega _{k,m}^{\star}=\frac{\varsigma _{k,m}}{1+1/\mathsf{SINR}_{k,m}\left(\bar{\mathbf{F}}_{m}^{\star}\right)}.
\end{equation}

\subsubsection{Fix $\boldsymbol{\omega }$ and Solve $\mathbf{\Theta}^{\star}$}
Then, by substituting the fixed $\boldsymbol{\omega }$ into \eqref{equ:P5}, the objective function can be transformed into a matrix form representation of $\boldsymbol{\theta }\triangleq\mathbf{\Theta }\mathbf{1}_{RN_{\mathrm{RIS}}}$, which is expressed as
\begin{equation}
f_5\left( \mathbf{\Theta } \right)=-\boldsymbol{\theta }^{\mathsf{T}}\mathbf{D}\boldsymbol{\theta }+2\mathfrak{R} \mathfrak{e} \left\{ \boldsymbol{\theta }^{\mathsf{T}}\tilde{\mathbf{d}} \right\} +\varepsilon 
,
\end{equation}
where 
\begin{subequations}
\begin{equation}
	\mathbf{D}\triangleq \sum_{k\in \mathcal{K}}{\sum_{m\in \mathcal{M}}{\sum_{j\in \mathcal{K}}{\mathbf{d}_{k,m,j}\mathbf{d}_{k,m,j}^{\mathsf{H}}}}},
\end{equation}	
\begin{equation}
	\tilde{\mathbf{d}}\triangleq \sum_{k\in \mathcal{K}}{\sum_{m\in \mathcal{M}}{\left( \varsigma _{k,m}\mathbf{d}_{k,m,k}-\sum_{j\in \mathcal{K}}{c_{k,m,j}^{*}\mathbf{d}_{k,m,j}} \right)}},
\end{equation}	
\begin{equation}
	\varepsilon \!\triangleq \!\!\sum_{k\in \mathcal{K}}\!{\sum_{m\in \mathcal{M}}\!{\!\!\left(\! 2\varsigma _{k,m}\mathfrak{Re} \!\left\{ c_{k,m,k} \right\} \!-\!\left| \omega _{k,m} \right|^2\!\sigma _{k,m}^{2}\!\!-\!\!\sum_{j\in \mathcal{K}}{\left| c_{k,m,j} \right|^2}\! \right)}},
\end{equation}	
\end{subequations}
where $\mathbf{d}_{k,m,j}\!\triangleq \!\sum_{a\in \mathcal{A}}{\omega _{k,m}^{*}\mathrm{diag}\!\left( \mathbf{u}_{k,m}^{\mathsf{H}} \right)\! \mathbf{T}_m\mathbf{G}_{a,m}\bar{\mathbf{F}}_{a,m}^{\star}\mathbf{w}_{a,m,j}}$
and $c_{k,m,j}\triangleq \sum_{a\in \mathcal{A}}{\omega _{k,m}^{*}\mathbf{h}_{\mathrm{dir},a,k,m}^{\mathsf{H}}\bar{\mathbf{F}}_{a,m}^{\star}\mathbf{w}_{a,m,j}}$. Therefore, the optimization of $\mathbf{\Theta}$ can be formulated as 
\begin{equation}\label{equ:P6}
\begin{split}
	(\mathrm{P}6)\,\, &\min_{\mathbf{\Theta }}\,\, f_6(\mathbf{\Theta })=\boldsymbol{\theta }^{\mathsf{T}}\mathbf{D}\boldsymbol{\theta }-2\mathfrak{R} \mathfrak{e} \left\{ \boldsymbol{\theta }^{\mathsf{T}}\tilde{\mathbf{d}} \right\}\\
	\,\,&\mathrm{s}.\mathrm{t}.\left| \theta _{r,n} \right|\in \mathcal{F} ,\forall r\in \mathcal{R} ,n=1,...,N_{\mathrm{RIS}}\\
\end{split}
\end{equation}
Similar to $\left( \mathrm{P}4 \right) $ in \eqref{equ:P4}, $\left( \mathrm{P}6 \right) $ can be solved by the PDS method.

\subsection{Precoding Design for TD of RIS $\mathbf{T}_m$}
Similar to the optimization of $\mathbf{\Theta }$, we exploit the MCQT again. By introducing auxiliary variables $\boldsymbol{\gamma }\triangleq\left[ \gamma_{1,1},\cdots ,\gamma _{K,1},\cdots ,\gamma _{K,M} \right]^{\mathsf{T}}$, the optimization subproblem of $\mathbf{\mathbf{T}_m}$ can be formulated as
\begin{equation}\label{equ:P7}
\begin{split}
	\left( \mathrm{P}7 \right) \,\,&\max_{\mathbf{T}_m,\boldsymbol{\gamma }} \,\,f_7\left( \mathbf{T}_m,\boldsymbol{\gamma } \right) =\sum_{k\in \mathcal{K}}{\sum_{m\in \mathcal{M}}{g_7\left( \mathbf{T}_m,\boldsymbol{\gamma } \right)}}
	\\
	\,\,&\mathrm{s}.\mathrm{t}.\left| t_{r,m,n}^{\mathrm{RIS}} \right|=1 ,\forall r\in \mathcal{R},m\in \mathcal{M},n=1,...,N_{\mathrm{RIS}},
\end{split}
\end{equation}
where
\begin{equation}\label{equ:g7}
\begin{split}
	g_7\left( \mathbf{T}_m,\boldsymbol{\gamma } \right) =&2\varsigma _{k,m}\Re \mathfrak{e} \left\{ \gamma _{k,m}^{*}\mathbf{h}_{k,m}^{\mathrm{H}}\overline{\mathbf{F}}_{m}^{\star}\mathbf{w}_{k,m} \right\}\\
	&-\left| \gamma _{k,m} \right|^2\left( \sum_{j\in \mathcal{K}}{\left| \mathbf{h}_{k,m}^{\mathrm{H}}\overline{\mathbf{F}}_{m}^{\star}\mathbf{w}_{j,m} \right|^2}+\sigma _{k,m}^{2} \right) .
\end{split}
\end{equation}
To solve subproblem $\left( \mathrm{P}7 \right)$, we update $ \mathbf{T}_m$ and $ \boldsymbol{\gamma }$ alternatively. 

\subsubsection{Fix $ \mathbf{T}_m$ and Solve $\boldsymbol{\gamma  }^{\star}$}
Firstly, similar to the update of $\mathbf{\Theta}$, the optimal $\boldsymbol{\gamma  }^{\star}$ can be obtained by solving $\frac{\partial f_7\left( \boldsymbol{\gamma } \right)}{\partial \gamma _{k,m}}=0$, given by
\begin{equation}\label{equ:gamma_*}
	\gamma _{k,m}^{\star}=\frac{\varsigma _{k,m}}{1+1/\mathsf{SINR}_{k,m}\left(\bar{\mathbf{F}}_{m}^{\star}\right)}.
\end{equation}

\subsubsection{Fix $\boldsymbol{\gamma  }$ and Solve $ \mathbf{T}_m^{\star}$}
Then, by substituting the fixed $\boldsymbol{\gamma}$ into \eqref{equ:P7} and defining $\boldsymbol{t}_m\triangleq\mathbf{T}_m\mathbf{1}_{RN_{\mathrm{RIS}}}$, the update of $\mathbf{T}_m$ can be reformulated as
\begin{equation}\label{equ:P8}
\begin{split}
	(\mathrm{P}8)\,\, &\min_{\boldsymbol{t}}\,\, f_8\left( \boldsymbol{t} \right) =\boldsymbol{t}^{\mathsf{T}}\mathbf{\Phi }\boldsymbol{t}-2\mathfrak{R} \mathfrak{e} \left\{ \boldsymbol{t}^{\mathsf{T}}\tilde{\boldsymbol{\phi}} \right\}   \\
	\,\,&\mathrm{s}.\mathrm{t}.\left| t_{r,m,n}^{\mathrm{RIS}} \right|=1 ,\forall r\in \mathcal{R},m\in \mathcal{M},n=1,...,N_{\mathrm{RIS}}\\
\end{split}
\end{equation}
where $\boldsymbol{t}\triangleq\left[ \boldsymbol{t}_{1}^{\mathsf{T}},\cdots ,\boldsymbol{t}_{M}^{\mathsf{T}} \right] ^{\mathsf{T}}$, $\mathbf{\Phi }\triangleq \mathrm{diag}\left\{ \mathbf{\Phi }_1,...,\mathbf{\Phi }_M \right\} $ and $\tilde{\boldsymbol{\phi}}\triangleq\left[ \tilde{\boldsymbol{\phi}}_{1}^{\mathsf{T}},\cdots ,\tilde{\boldsymbol{\phi}}_{M}^{\mathsf{T}} \right] ^{\mathsf{T}}$. Besides,
\begin{subequations}
	\begin{equation}
		\mathbf{\Phi }_m\triangleq \sum_{k\in \mathcal{K}}{\sum_{j\in \mathcal{K}}{\bar{\mathbf{d}}_{k,m,j}\bar{\mathbf{d}}_{k,m,j}^{\mathsf{H}}}},
	\end{equation}	
	\begin{equation}
		\tilde{\boldsymbol{\phi}}_m\triangleq \sum_{k\in \mathcal{K}}{\left( \varsigma _{k,m}\bar{\mathbf{d}}_{k,m,k}-\sum_{j\in \mathcal{K}}{\bar{c}_{k,m,j}^{*}\bar{\mathbf{d}}_{k,m,j}} \right)},
	\end{equation}	
\end{subequations}
with $\bar{\mathbf{d}}_{k,m,j}\!\triangleq \!\sum_{a\in \mathcal{A}}{\gamma _{k,m}^{*}\mathrm{diag}\!\left( \mathbf{u}_{k,m}^{\mathsf{H}}\mathbf{\Theta }_k \right) \!\mathbf{G}_{a,m}\bar{\mathbf{F}}_{a,m}^{\star}\mathbf{w}_{a,m,j}}$ and $\bar{c}_{k,m,j}\triangleq \sum_{a\in \mathcal{A}}{\gamma _{k,m}^{*}\mathbf{h}_{\mathrm{dir},a,k,m}^{\mathsf{H}}\bar{\mathbf{F}}_{a,m}^{\star}\mathbf{w}_{a,m,j}}$. Therefore, the convex problem $	(\mathrm{P}8)$ formulated in \eqref{equ:P8} can be solved by PDS algorithm.

\section{Numerical Results and Discussion}\label{sec:Numerical Results and Discussion}
In this section, we present simulation results by analyzing convergence and the impact of key system parameters to demonstrate the effectiveness and robustness of the proposed optimization framework.

\subsection{Simulation Setup}
\begin{figure}[!t]
	\centering
	\includegraphics[width=0.5\textwidth]{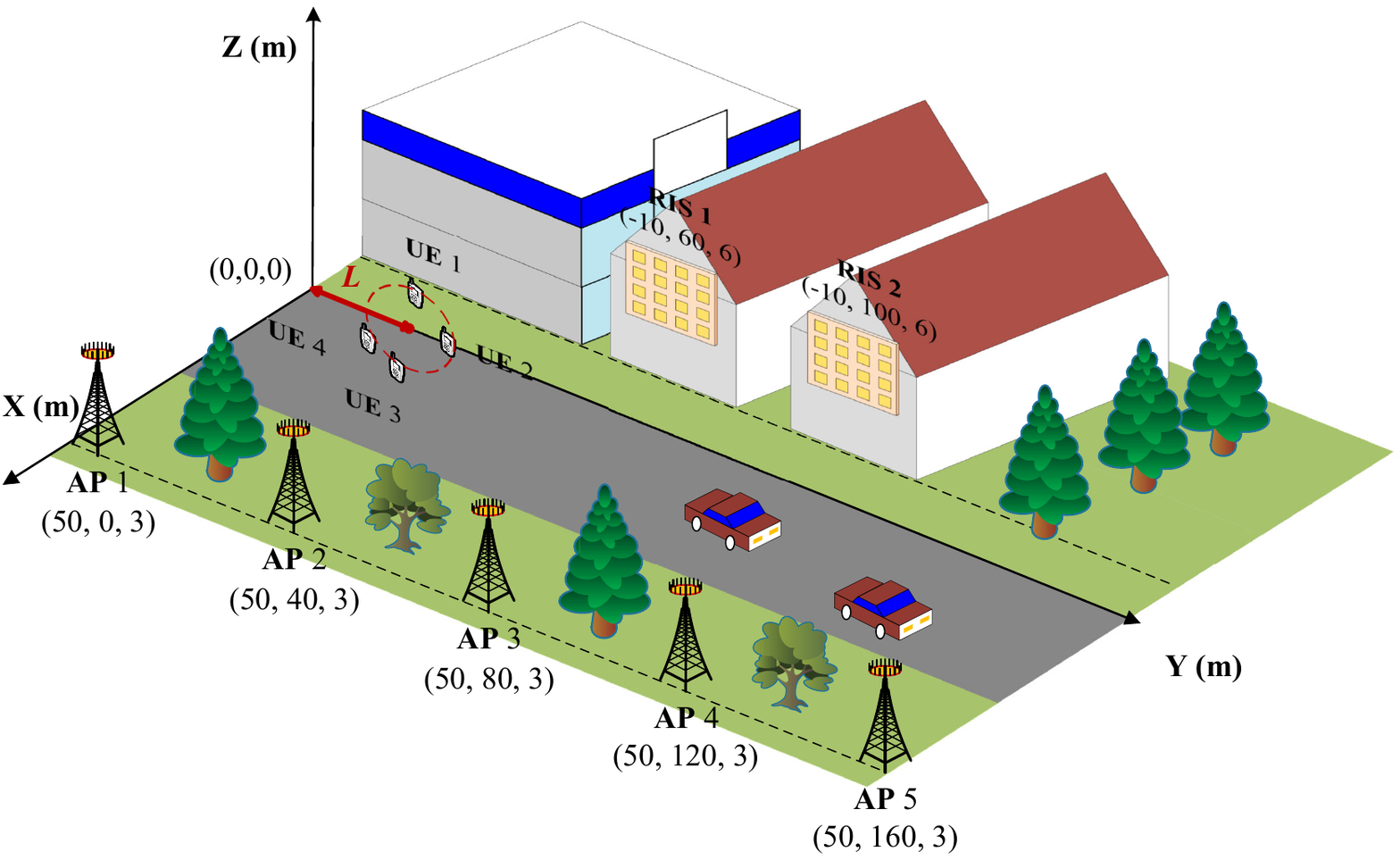}
	\caption{3D-simulation scenario of RIS-assisted THz cell-free mMIMO system.}
	\label{fig:Plot_Model}
\end{figure}

For the simulation, we consider a 3D scenario as shown in Fig. \ref{fig:Plot_Model}. In this considered scenario, a cell-free network with $A=5$ APs serve $K=4$ users with the assistance of $R=2$ RISs \cite{sun2024geometric}. Assume that the four users are randomly distributed around the central point $(L,0)$.
For simplicity and without loss of generality, the remaining parameters are set as $f_c= 100$ GHz, $B = 10$ GHz, $N_\mathrm{Tx}= 16$, $N_\mathrm{X}=N_\mathrm{Y}= 8$, $N_\mathrm{TD}=16$, $M= 8$, $\sigma^2=-110$ dBm, $\left\{ P_{a,\max} \right\} _{a\in \mathcal{A}}=0$ dBm.  Besides, both large and small scale fading parameters adopt the same setting as \cite{zhang2021joint}. Moreover, $\mathbf{w}$, $\mathbf{\Theta}$ and $\mathbf{T}_m$ is initialized by random values satisfying the constraint in \eqref{equ:P1}.

In the following figures, the curves ``proposed", ``proposed 1-bit" and ``proposed 2-bit" represent the proposed optimization framework with ideal RIS, non-ideal 1-bit phase RIS and 2-bit phase RIS, respectively. Besides the curves ``without RIS" and ``without TD" represent the TD-equipped cell-free networks without RISs and RIS-assisted cell-free networks without TDs, respectively. 

\subsection{Convergence}
\begin{figure}[!t]
	\centering
	\includegraphics[width=0.5\textwidth]{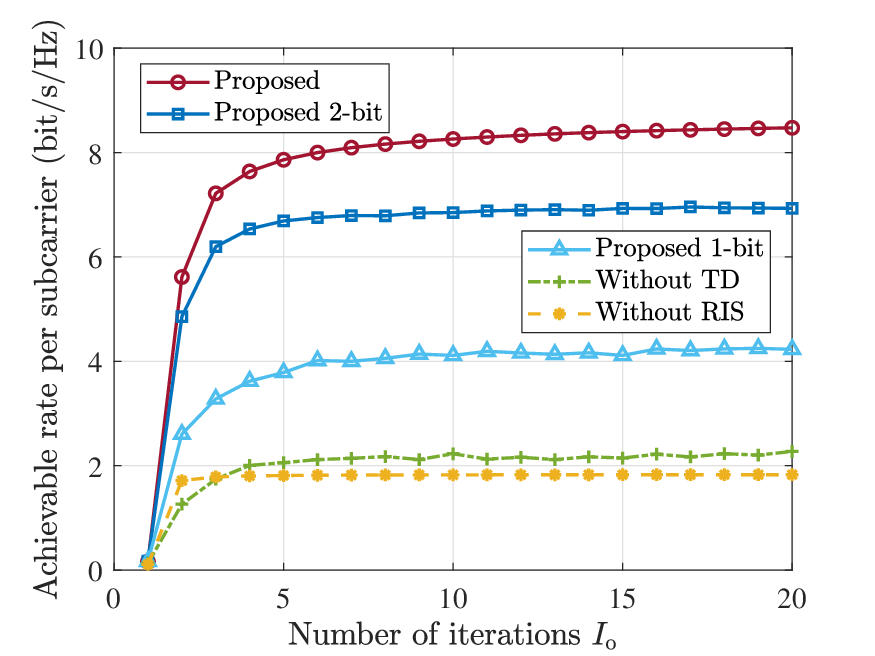}
	\caption{ASR performance against the iteration number $N_{\mathrm{iter}}$.}
	\label{fig:ASR_N}
\end{figure}

To illustrate the convergence of the proposed optimization framework, the average sum-rate (ASR) at each SC versus the number of iterations $N_{\mathrm{iter}}$ is shown in Fig. \ref{fig:ASR_N}. It can be observed that the proposed framework and low-resolution RIS case (both 1-bit and 2-bit phase RIS) can converge within 20 iterations and 10 iterations, respectively. Compared with the proposed framework, the ``without RIS" case and ``without TD" case can converge within 5 iterations, this is because these cases have no need to address the precoding of RISs or TDs. Moreover, it can be observed that the proposed optimization framework even with low-resolution RISs can achieve superior ASR performance compared with the ``without RIS" case and ``without TD" case, which demonstrate the effectiveness of the proposed optimization framework.

\subsection{Robustness}
\begin{figure}[!t]
	\centering
	\includegraphics[width=0.5\textwidth]{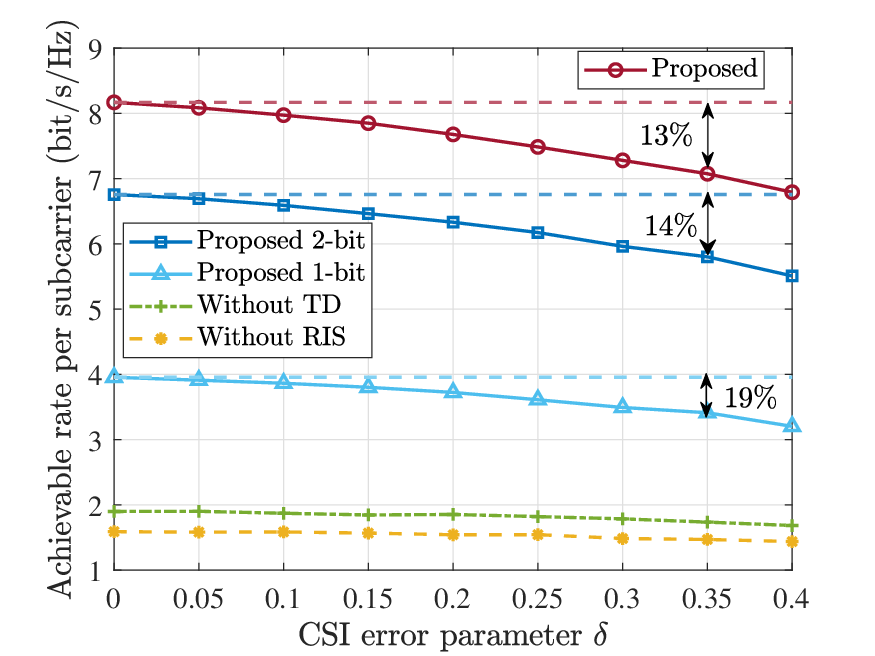}
	\caption{ASR performance against the CSI error parameter $\delta$.}
	\label{fig:ASR_e}
\end{figure}

Since channel estimation is a challenging work in RIS-assisted THz cell-free systems, we employ practical channel estimation error to validate the robustness of the proposed optimization framework. Assuming the practical estimated channel $\hat{h}$ is modeled as \cite{ubaidulla2011relay}
\begin{equation}
	\hat{h}=h+e ,
\end{equation}
where $h$ is the ideal channel and $e \sim \mathcal{C} N\left(0, \sigma_e^2\right)$ denotes the Gaussian estimation error. The error power satisfies $\sigma_e^2 \triangleq \delta|h|^2$, wherein the error-channel gain power ratio $\delta$ is defined as the CSI error parameter. Then, the ASR versus CSI error parameter is shown in Fig. \ref{fig:ASR_e}. It can be observed that the performance loss of all cases grows as $\delta$ increases. In particular, for the proposed optimization framework including non-ideal RIS case (both 1-bit and 2-bit phase RIS), compared with perfect CSI (i.e., $\delta=0$), the ASR of the three cases suffers a loss of around $13\%$, $14\%$, $19\%$, respectively, when the estimation error reaches $35\%$ of channel gain (i.e., $\delta=0.35$). Furthermore, the ASR performance of all these three cases even with high level estimation error (e.g., $\delta=0.4$) outperform the ``without RIS" case and ``without TD" case with perfect CSI (i.e., $\delta=0$). Therefore, the proposed optimization framework shows strong tolerance of imperfect CSI, which demonstrates its robustness against estimation error.

\subsection{Impact of Key System Parameters}
To futher demonstrate the effectiveness of the proposed optimization framework, we discuss the performance under different system parameters. As observed in Fig. \ref{fig:ASR_N}, we set $N_{\mathrm{iter}}=20$ to ensure the convergence of all algorithms. 
\subsubsection{ASR against AP Transmit Power}
\begin{figure}[!t]
	\centering
	\includegraphics[width=0.5\textwidth]{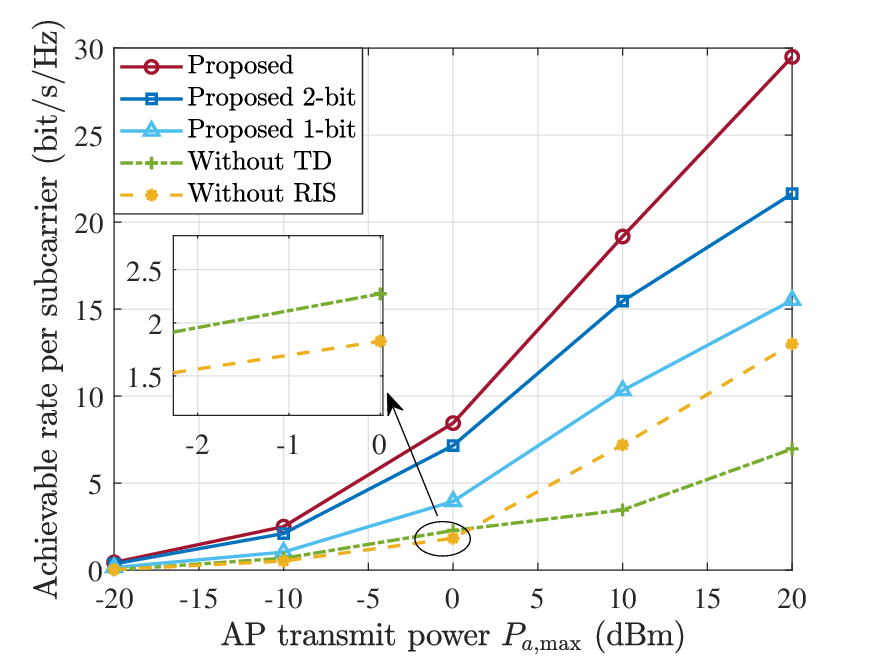}
	\caption{ASR performance against the AP transmit power $P_{a,\max}$.}
	\label{fig:ASR_P}
\end{figure}

Assuming $K=4$ users randomly distributed around the central point where $L=40$ m, the ASR versus the AP transmit power is depicted in Fig. \ref{fig:ASR_P}. It is shown that as the AP transmit power increases, the ASR in all case improve rapidly and the performance gaps among these cases become larger. Besides, the proposed optimization framework even employed with non-ideal RIS outperforms the other two case in all considered power regions. This indicates that the deployment of RISs and TDs can efficiently assist signal transmission and mitigate beam split effect. In addition, when the AP transmit power is insufficient (e.g., less than $0$ dBm), ``without RIS" case and ``without TD" case achieve similar ASR performance. However, while the AP transmit power is moderate (e.g., larger than $10$ dBm), the performance of ``without RIS" case is better than ``without TD" case. This is because under high AP transmit power, the signal transmission depends more on AP-UE direct links rather than RIS-assisted links.

\subsubsection{ASR against UE number}
\begin{figure}[!t]
	\centering
	\includegraphics[width=0.5\textwidth]{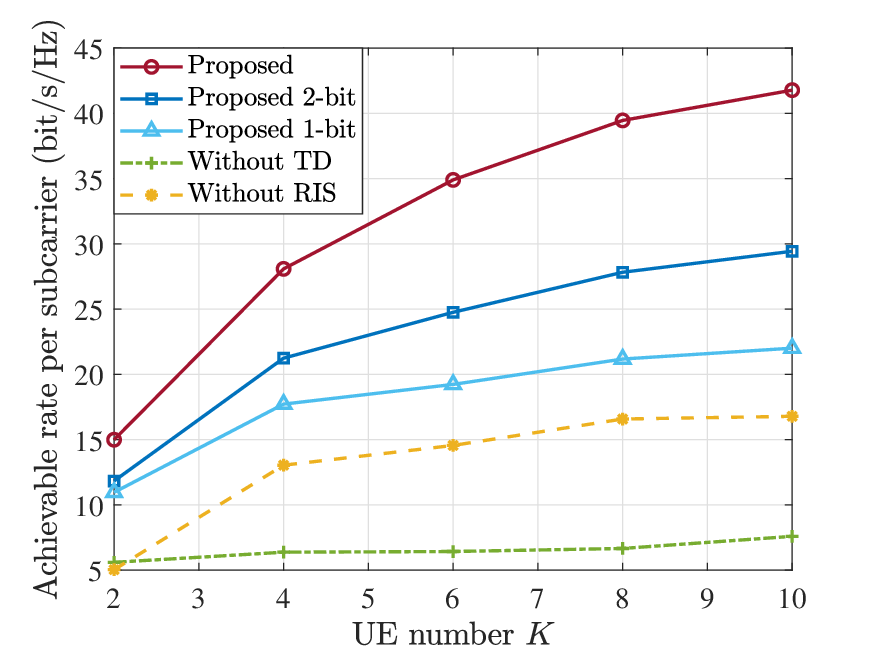}
	\caption{ASR performance against the UE number $K$.}
	\label{fig:ASR_K}
\end{figure}
Adopting the same parameters as above and setting the AP transmitted power as $P_{a,\max}=20$ dBm, the ASR versus the UE number $K$ is presented in Fig. \ref{fig:ASR_P}. We can observe that, the proposed framework even equipped with non-ideal RISs can achieve superior ASR performance compared with ``without RIS" case and ``without TD" case. Moreover, the ASR performance of all the five algorithms improves as UE number increases. However, with the increase in UE number, the growth extent of ASR diminishes. This is because the increased number of UEs introduces more severe inter-user interference, which limits the further improvement of system performance.

\subsubsection{ASR against UE Location}
\begin{figure}[!t]
	\centering
	\includegraphics[width=0.5\textwidth]{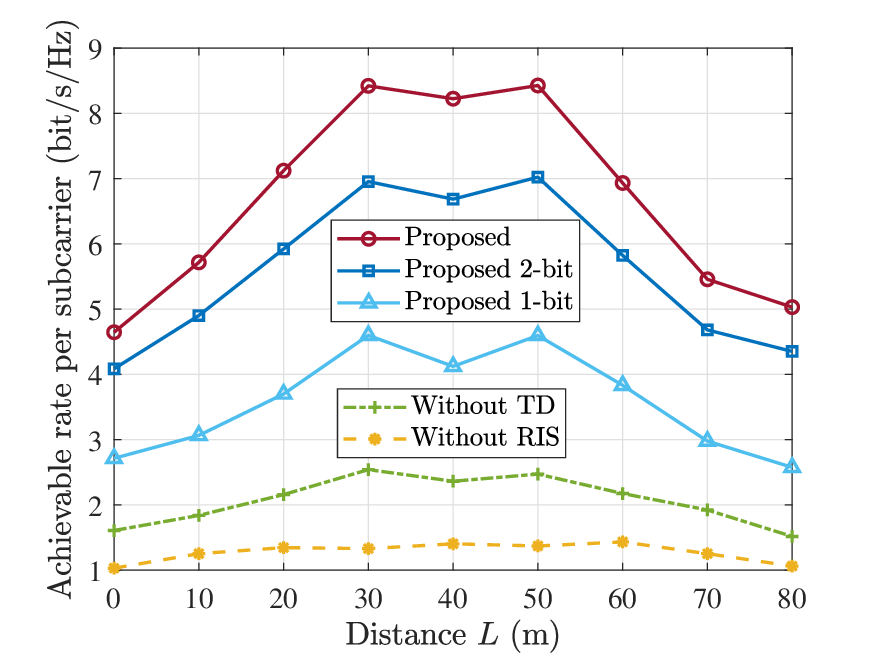}
	\caption{ASR performance against the distance $L$.}
	\label{fig:ASR_L}
\end{figure}
By fixing $P_{a,\max}=0$ dBm and $K=4$, we plot the ASR versus the distance $L$ of central point in Fig. \ref{fig:ASR_L}. It is shown that for the ``proposed",``proposed 1-bit" and ``proposed 2-bit" case which follow the proposed optimization framework, the ASR occurs obvious peaks when the users approach any of the two RISs. This is because the RISs can reflect  stronger signal towards users. Furthermore, the proposed framework even employing low-resolution RISs can achieve better ASR performance compared with the ``without RIS" case and ``without TD" case. This indicates that both RISs and TDs contribute to performance improvement.

\section{Conclusion}\label{sec:Conclusion}
In this paper, we investigated joint precoding framework for RIS-assisted wideband THz cell-free mMIMO systems. To overcome beam split effect, we introduced additional TD layers at both APs and RISs. Based on AO framework, we proposed a joint precoding design to solve the TDs of APs, baseband beamformers, phase shifts and TDs of RISs, respectively. Simulation results demonstrate that the introduction of RISs and TDs can effectively mitigate beam split effect and improve network capacity. The proposed joint precoding design is efficient under various condition and robust to estimation error, and it is suitable while employing low-resolution RISs.


\bibliographystyle{IEEEtran}
\bibliography{./bibtex/IEEEabrv,./bibtex/IEEEreference}


\end{document}